\documentclass[%
 reprint,
 superscriptaddress,
 amsmath,amssymb,
 aps,
 prl,
floatfix,
longbibliography 
]{revtex4-2}

\usepackage[pdftex]{graphicx} \graphicspath{{}}
\usepackage{float}
\usepackage{dcolumn}
\usepackage{bm}
\usepackage[pdftex,colorlinks=true]{hyperref}
\hypersetup{
  colorlinks=true,
  linkcolor=blue,
  urlcolor=cyan,
}

\newcommand{\eqn}[1]{\begin{equation} #1 \end{equation}}
\newcommand{\eqa}[1]{\begin{align} #1 \end{align}}

\usepackage{color}
\definecolor{ForestGreen}{RGB}{34, 139, 34}

\newcommand{\amcr}[1]{\textsf{\color{red}(#1)}}

\newcommand{\nn}{\nonumber}

\newcommand{\mH}{\mathcal{H}}

\newcommand{\pd}{\partial}
\newcommand{\bS}{\boldsymbol{S}}

\DeclareMathOperator{\sech}{sech}

\newcommand{\sectionn}[1]{\textit{#1---}}

\begin{document}

\preprint{APS/123-QED}

\title{Long-lived Solitons and Their Signatures in the Classical Heisenberg Chain}

\author{Adam J. McRoberts}
\affiliation{Max Planck Institute for the Physics of Complex Systems, N\"{o}thnitzer Str. 38, 01187 Dresden, Germany}

\author{Thomas Bilitewski}
 \affiliation{Max Planck Institute for the Physics of Complex Systems, N\"{o}thnitzer Str. 38, 01187 Dresden, Germany}
 \affiliation{JILA, NIST, and Department of Physics, University of Colorado, Boulder, CO 80309, USA}
 \affiliation{Center for Theory of Quantum Matter, University of Colorado, Boulder, CO 80309, USA}
 
\author{Masudul Haque}
 \affiliation{Max Planck Institute for the Physics of Complex Systems, N\"{o}thnitzer Str. 38, 01187 Dresden, Germany}
 \affiliation{Department of Theoretical Physics, Maynooth University, Co. Kildare, Ireland}
 \affiliation{Institut f\"ur Theoretische Physik, Technische Universit\"at Dresden, 01062 Dresden, Germany}

\author{Roderich Moessner}
 \affiliation{Max Planck Institute for the Physics of Complex Systems, N\"{o}thnitzer Str. 38, 01187 Dresden, Germany}

 
\date{\today}

\begin{abstract}
\noindent

Motivated by the KPZ scaling recently observed in the classical ferromagnetic Heisenberg chain, we investigate the role of solitonic excitations in this model.  We find that the Heisenberg chain, although well-known to be non-integrable, supports a two-parameter family of long-lived solitons.  We connect these to the exact soliton solutions of the integrable Ishimori chain with $\log(1+ S_i\cdot S_j)$ interactions.  We explicitly construct infinitely long-lived stationary solitons, and provide an adiabatic construction procedure for moving soliton solutions, which shows that Ishimori solitons have a long-lived Heisenberg counterpart when they are not too narrow and not too fast-moving. 
Finally, we demonstrate  their presence in thermal states of the Heisenberg chain, even when the typical soliton width is larger than the spin correlation length, and argue that these excitations likely underlie the KPZ scaling.
\end{abstract}

\maketitle

\sectionn{Introduction}
There has been renewed interest in understanding the long-time dynamics of classical many-body systems, in particular regarding the scope of anomalous, non-diffusive, transport. A paradigmatic phenomenon is  Kardar-Parisi-Zhang (KPZ) scaling \cite{KPZ_1986}, associated with (generalised) hydrodynamics \cite{bertini2016transport,bulchandani2017solvable,doyon2018soliton,doyon2019lecture,castroalvaredo2016emergent,bulchandani2018bethe,Schemmer2019,das2020nonlinear,Doyon_2018,doyon2018soliton,doyon2019lecture} and integrability \cite{Spohn_2016,Spohn_2014a,Spohn_2015,DeNardis2018,Sasamoto_2018,das2019kardar,das2020nonlinear,Roy2022KPZ,mcroberts2022anomalous,Lepri_Politi_PRL2020,Spohn_JStatPhys2014,Dhar_Spohn_FPU_PRE2014,Oleksandr2019,Krajnik2020,bulchandani2020KPZ}. Recent theoretical developments have identified integrability and non-abelian symmetry as key ingredients for KPZ physics \cite{prosen2013macroscopic,Krajnik2020,DeNardis_2021,Ilievski_2021,dupont2020universal,das2019kardar,bulchandani2020KPZ}.

Indeed, KPZ scaling is now established  \cite{das2019kardar,das2020nonlinear} in the integrable Ishimori chain \cite{ishimori1982integrable}, also known as the integrable lattice-Landau-Lifshitz model. Intriguingly, the simple {\it non-integrable} nearest-neighbour classical Heisenberg chain was also found to host a long-lived regime of KPZ scaling at low temperature  \cite{mcroberts2022anomalous}, and it was subsequently noted that KPZ scaling in the Ishimori chain persists under spin-symmetry preserving perturbations \cite{Roy2022KPZ}. 

\begin{figure}[ht!]
    \centering
    \includegraphics{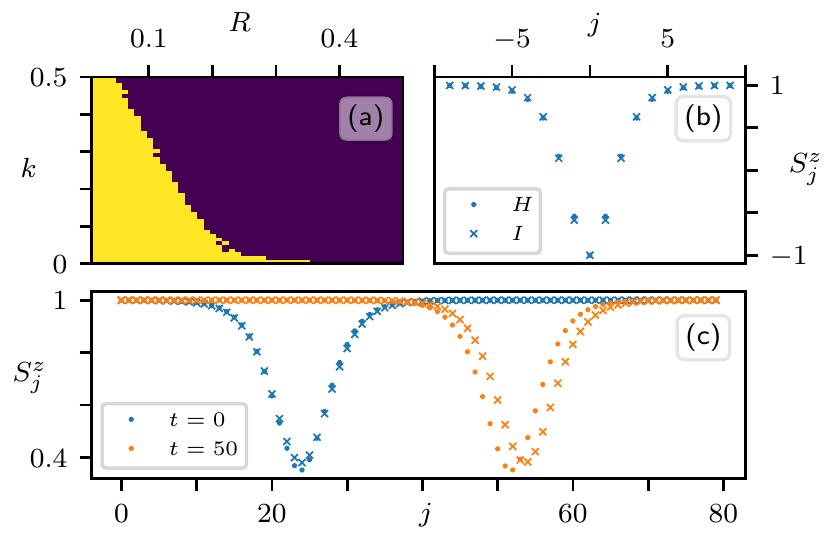}
    \caption{Solitons in the classical Heisenberg chain. (a)  parameter space of Ishimori solitons for which we found an adiabatically connected soliton in the Heisenberg chain. (b) \& (c)  comparisons between adiabatically connected solitons (only  $z-$component shown) in the Heisenberg ($H$, $\bullet$) and Ishimori ($I$, $\times$) chains. The solitons in (b) are stationary, $(R = 0.25, k = 0)$; and in (c) they move, $(R = 0.1, k = 0.15)$. The Heisenberg soliton moves at a slower velocity, but both preserve their initial profile. 
    }
    \label{fig:solitons}
\end{figure}
Whilst the classical Heisenberg spin chain is a widely studied system and  a paradigmatic model of magnetism, it remains far from completely understood. For example, predictions of its hydrodynamics have involved ordinary diffusion \cite{gerling1989comment, gerling1990time, bohm1993comment, srivastava1994spin, oganesyan2009energy, bagchi2013spin, glorioso2020hydrodynamics} or different forms of anomalous behaviour \cite{muller1988anomalous, de1992breakdown, de1993alcantara, de2020universality}.  Our recent observation of KPZ behaviour up to enormously large  scales \cite{mcroberts2022anomalous} thus raises the question: does the Heisenberg chain exhibit properties ordinarily associated with integrability? 
In particular, one might wonder if this phenomenology can be related to magnon dynamics or the existence of solitons, thought to be crucial for KPZ behaviour both in quantum  \cite{bulchandani2020KPZ,de2020superdiffusion,gopalakrishnan2019kinetic,DeNardis_2021,ljubotina2019KardarParisiZhang} and classical integrable 1D spin systems \cite{bulchandani2020KPZ,das2019kardar,das2020nonlinear,Roy2022KPZ}.  

In this letter, we demonstrate the existence of long-lived solitons in the classical Heisenberg chain. The appellation soliton is justified by an explicit continuous connection to those of the Ishimori chain via an interpolating Hamiltonian.
We provide a direct construction of stable ({\it infinitely} long-lived) \textit{stationary} isolated solitons, as well as an adiabatic construction of \textit{moving} solitons. A central result is the existence of a family of solitons which are stable over a broad parameter regime (see Fig.~\ref{fig:solitons}(a)). This is, \textit{a priori}, very surprising for a chain so far believed to be essentially generic. %
Beyond the isolated solitons, we study two-soliton scattering and observe behaviour quite analogous to that of the integrable model. Finally, for low-temperature thermal states, we show that solitons are present and can be individually identified even when their density is high. %
Taken together, these observations provide a physical basis for the robust KPZ scaling observed at low temperatures in the Heisenberg chain.


\sectionn{Models}
The classical Heisenberg Hamiltonian is
\eqn{
\mH = -J \sum_i \left( \bS_i \cdot \bS_{i+1} - 1 \right),
\label{H_Heisenberg}
}
where $\mathbf{S}_i$ are classical $O(3)$ vectors at sites $i$ of a chain, with nearest-neighbour ferromagnetic interaction strength $J$. 

The integrable \cite{ishimori1982integrable,theodorakopoulos1995nontopological,theodorakopoulos1988semiclassical, book_FaddeevTakhtajan_1987_HamiltonianMethods, Sklyanin_JournSovMath1988, Sklyanin_FuncAnal1982, Prosen_Zunovic_PRL2013} Ishimori Hamiltonian,
\eqn{
\mH = -2J \sum_i \log \left(\frac{1 + \bS_i \cdot \bS_{i+1}}{2} \right),
\label{H_Ishimori}
}
possesses an extensive set of locally conserved charges, besides 
energy and magnetisation, such as the torsion
\eqn{
\tau_i = \frac{\bS_i \cdot (\bS_{i + 1} \times \bS_{i - 1})}{(1 + \bS_i \cdot \bS_{i + 1})(1 + \bS_i \cdot \bS_{i - 1})}.
\label{tau}
}


We interpolate smoothly between the  chains,
\eqn{
\mH = - 2J \gamma^{-1} \sum_i \log \left(1 + \gamma \frac{\bS_i \cdot \bS_{i+1} - 1}{2} \right),
\label{H_gamma}
}
with the Ishimori chain corresponding to $\gamma = 1$, and the Heisenberg chain to the limit $\gamma \rightarrow 0$,  preserving $SO(3)$ symmetry throughout. We set $J = 1$ in the following.

The classical equations of motion follow from
\eqn{
\dot{\bS_i} = \frac{\pd \mH}{\pd \bS_i} \times \bS_i,
}
from which we obtain the dynamics of Eq.~(\ref{H_gamma}):
\eqn{
\dot{\bS_i} = 2 \bS_i \times \left( \frac{\bS_{i-1}}{2 - \gamma + \gamma \bS_i \cdot \bS_{i-1}} + \frac{\bS_{i+1}}{2 - \gamma + \gamma \bS_i \cdot \bS_{i+1}} \right).
\label{eom}
}

\sectionn{One-Soliton Solutions}
In  the Ishimori chain \cite{ishimori1982integrable}, these are indexed by two physical parameters: an inverse-width $R \in (0, \infty)$ and a wavenumber $k \in [-\pi/2, \pi/2)$, see SI \cite{supplemental} for explicit expressions and their properties. 
The Heisenberg chain (\ref{H_Heisenberg}), by contrast, is not integrable. We next provide exact (though not closed-form) expressions for stationary solitons in the form of an (implicit) solution to the non-linear equations of motion of the Heisenberg model. 

For this, we use canonical co-ordinates, $z_i = S^z_i$, $\phi_i = \arctan(S^y_i/S^x_i)$. 
Our ansatz is based on the structure of the stationary ($k = 0$) Ishimori solitons. We assume  (i) stationarity of the z-components, i.e., $\dot{z}_i = 0,\, \forall i $
, (ii) spatially uniform azimuthal angles $\phi_i$ (except for a discontinuity of $\pi$ across the centre), and (iii) a uniform rotation frequency of the in-plane spin-components, i.e., $\phi_i(t) = \phi_i(0) + \omega t, \; \forall i$. 
This ansatz reduces the equations of motion to a set of consistency equations for the $z_i$,
\eqa{
\dot{\phi}_i = \omega = &J \frac{z_i}{\sqrt{1 - z_i^2}} \left(\sqrt{1 - z_{i+1}^2} + \sqrt{1 - z_{i-1}^2} \right) \nn \\
&-J \left(z_{i+1} + z_{i-1}\right),
\label{sol_theta}
}
which, for a chosen frequency $\omega$, may be solved numerically to arbitrary precision \cite{supplemental}.

This yields stable stationary solitons of arbitrary width ($\sim 1/R$), implying that the existence diagram in Fig.~\ref{fig:solitons}(a) extends to infinity on the x-axis. An example of a soliton obtained from the solution of these equations is shown in Fig.~\ref{fig:solitons}(b). This constitutes the first (to our knowledge) exact soliton in the Heisenberg model. 

\sectionn{Adiabatic connection}
We next connect these stationary solitons to those in the Ishimori chain by continuously tuning the interpolating Hamiltonian (\ref{H_gamma}) between the two via 
a $C^{\infty}$-smooth interpolation,
\eqn{
\gamma(t) =
\begin{cases}
1 - \frac{e^{-t_A/t}}{e^{-t_A/t} + e^{-t_A/(t_A - t)}} & 0 < t < t_A \\
0 & t \geq t_A
\end{cases}
\label{adiabatic}
}
from $\gamma = 1$ at $t=0$ to $\gamma = 0$ at some long adiabatic time $t_A$. We evolve an initial Ishimori soliton $(R, k)$ under the dynamics of Eq.~(\ref{H_gamma}), with this time-dependent $\gamma(t)$ given by Eq.~(\ref{adiabatic}), for some adiabatic time $t_A$; we then evolve up to some later time $t_f$ under the Heisenberg dynamics (\ref{H_Heisenberg}). 

This continuously transforms stationary solitons of the Ishimori chain into stationary solitons of the Heisenberg chain with the same magnetisation.

\begin{figure}
    \centering
    \includegraphics{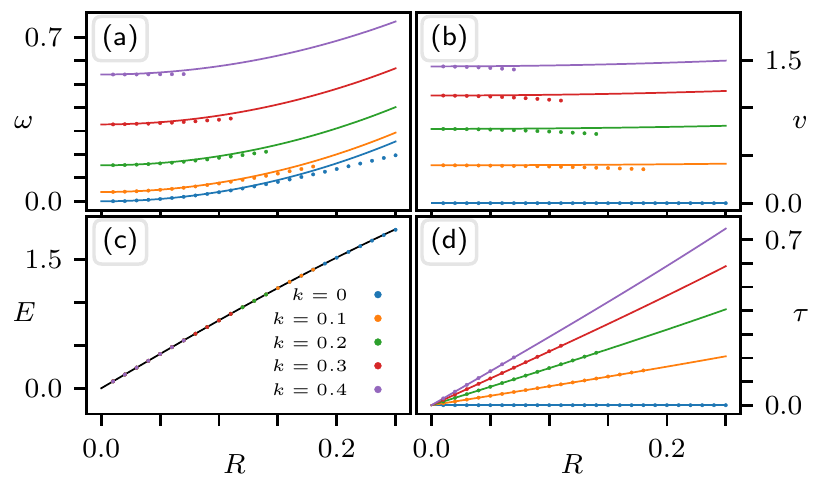}
    \caption{Physical properties of the Ishimori solitons (solid lines), compared to Heisenberg solitons (dotted lines -- up to the boundary of the existence diagram, Fig.~\ref{fig:solitons}(a)), shown as a function of inverse-width $R$, for various values of the wavenumber $k$. (a), (b), (c), and (d), show, resp., the internal frequency $\omega$, the velocity $v$, the energy $E$ (measured by the Heisenberg Hamiltonian) and the torsion $\tau$.
    }
    \label{fig:renormalisation}
\end{figure}

\sectionn{Moving solitons}
As the connection between the stationary solitons of the two models does not guarantee the existence of moving soliton solutions of the Heisenberg chain, we next use our adiabatic procedure to extend  the `existence diagram' in Fig~\ref{fig:solitons}(a) to finite $k$. We consider a resultant state a soliton of the Heisenberg chain if the following conditions are satisfied: (i) there is, for all times, a unique local minimum of $z_i(t)$; (ii) for $t > t_A$, the torsion $\tau = \sum_i \tau_i$ is constant in time; (iii) the unique local minimum propagates with a constant velocity. These conditions are examined in more detail in \cite{supplemental}. An example of a thus constructed moving soliton is shown in Fig.~\ref{fig:solitons}(c), and compared to the original Ishimori soliton.

The resulting existence diagram (Fig.~\ref{fig:solitons}(a)), shows  that the solitons are stable in the Heisenberg model over a remarkably large range of parameters $(R,k)$, with the narrow solitons apparently becoming unstable first with increasing velocity ($\sim k$). 

We find no indication of a finite lifetime of the single soliton states which are stable under this adiabatic procedure. Moreover, the torsion -- generally not a conserved quantity of the Heisenberg chain -- is conserved in these states. Whilst stationary solutions of non-linear classical equations of motion are well known (see e.g. \cite{FLACH20081}), stable moving solitons are not generally expected to exist. 

Next we consider how the properties of Ishimori solitons are modified in the Heisenberg model. Fig.~\ref{fig:renormalisation} shows that the internal frequency (the frequency with which the in-plane spin-components rotate) and velocity of a Heisenberg soliton are suppressed. The energy (measured in both cases by the Heisenberg Hamiltonian) is only slightly reduced -- whilst the torsion is very slightly higher for the Heisenberg solitons (see also Fig.~\ref{fig:stability}, \cite{supplemental}). Overall, we note a remarkable similarity between the one-soliton properties in the Ishimori and Heisenberg chain.

\sectionn{Two-soliton scattering}
We now turn to interactions between the solitons.
To set the stage, we  briefly recall scattering in the Ishimori chain.
As a fully integrable model, interactions are completely described by the two-soliton phase-shifts, even for thermal multi-soliton states \cite{theodorakopoulos1995nontopological,theodorakopoulos1988semiclassical}. When two 
solitons collide, the asymptotic result (compared to two separate one-soliton solutions) is unchanged, except that the solitons are displaced by a so-called phase-shift \cite{theodorakopoulos1995nontopological}:
\eqa{
\Delta(&R, k; R', k') = \mathrm{sgn}(v(R, k) - v(R', k')) \nn \\ 
&\times \frac{1}{2R} \log \left[\frac{\cosh(2(R + R')) - \cos(2(k - k'))}{\cosh(2(R - R')) - \cos(2(k - k'))}\right].
\label{phase_shift}
}
experienced by the soliton $(R, k)$, due to a collision with the soliton $(R', k')$.

\begin{figure}
    \centering
    \includegraphics{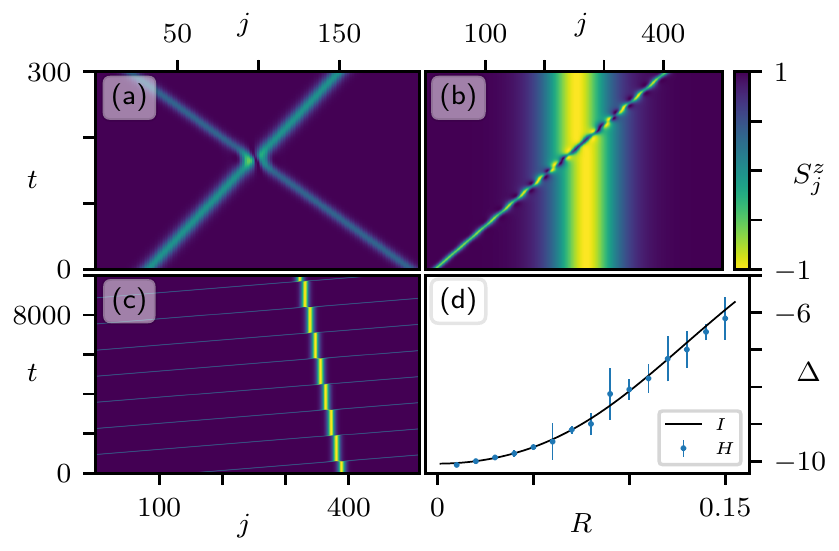}
    \caption{Soliton scattering in the Heisenberg chain. Colour-scale shows the $z$-components, and is the same for (a), (b), and (c). (a) Single scattering event between two solitons with parameters $(R, k) = (0.1, 0.1)$ and $(R', k') = (0.1, -0.15)$. (b) Screening of the magnetisation transported by a narrower soliton as it moves through a wider soliton. (c) Repeated scattering of two solitons [$(R, k) = (0.1, 0.1)$ and $(R', k') = (0.1, 0)$] under periodic boundary conditions. (d) Comparison of the scattering phase-shift $\Delta(R, k = 0; 0.1, 0.1)$ in the Ishimori chain (solid line) and in the Heisenberg chain, for stationary target solitons. Phase-shifts in the Heisenberg chain obtained by averaging over 10 scattering events, cf. (c), and over the relative phases of the solitons -- the error bars are the standard deviation w.r.t. the relative phases.}
    \label{fig:scattering}
\end{figure}

Fig.~\ref{fig:scattering} displays the scattering of two Heisenberg solitons. The solitons survive scattering essentially unchanged (Fig.~\ref{fig:scattering}(a)), akin to the fully integrable model. Whilst the collisions do leave the solitons unchanged asymptotically, the magnetisation of a moving soliton is 
`screened' during the collision with a larger soliton as seen in Fig.~\ref{fig:scattering}(b). Importantly, solitons survive multiple collisions (Fig.~\ref{fig:scattering}(c)), with the change to their trajectories apparently given by simple consecutive phase-shifts. 

There exist, nonetheless, some important differences between the Heisenberg and Ishimori cases. First, absent integrability, scattering in the Heisenberg chain is not expected to be perfectly lossless. Indeed, there is a very small amount of radiation emitted during the collision (approx. $\delta S^z \sim 10^{-6}$ in magnitude) in Fig.~\ref{fig:scattering}(a). Second, scattering from narrow solitons at small $k$ (where the existence diagram is wider in $R$) can emit significant amounts of radiation, although, curiously, the modified solitons that emerge appear to be stable to subsequent collisions. 

In addition, the phase-shift $\Delta$ appears not to depend only on the soliton parameters $R, k, R', k'$. We extract the phase-shifts in Fig.~\ref{fig:scattering}(d) by averaging over 10 scattering events. They are also averaged over the relative phase (azimuthal angles) of the solitons at the moment of collision. In the integrable case, this has no effect -- in the Heisenberg case, however, in particular for larger $R$, the phase-shift depends on the relative phase (see Fig.~\ref{fig:scatter3}).  We discuss these points in more detail in \cite{supplemental}.

Despite these differences, however, we emphasise that collisions over a large parameter regime in the Heisenberg model strongly resemble the scattering in the Ishimori case. Importantly, whilst the phase-shift $\Delta$ acquires some fluctuations, the velocities of the solitons remain unaffected by the collisions.

\sectionn{Solitons in thermal states}
Whilst the Heisenberg chain supports solitons as stable solitary waves, which suffer only very weak dissipation in scattering events, the imperfect nature of the scattering 
implies the existence of a thermal timescale on which they eventually decay. The question arises, then, as to what extent these solitons exert their influence on the hydrodynamics and transport properties: thermal states are not in any sense a dilute soliton gas, and it would not be unreasonable to expect solitons to experience so many scattering events that, unprotected by integrability, they collapse too swiftly to generate a discernible superdiffusive contribution to the transport of spin or energy.

\begin{figure}
    \centering
    \includegraphics{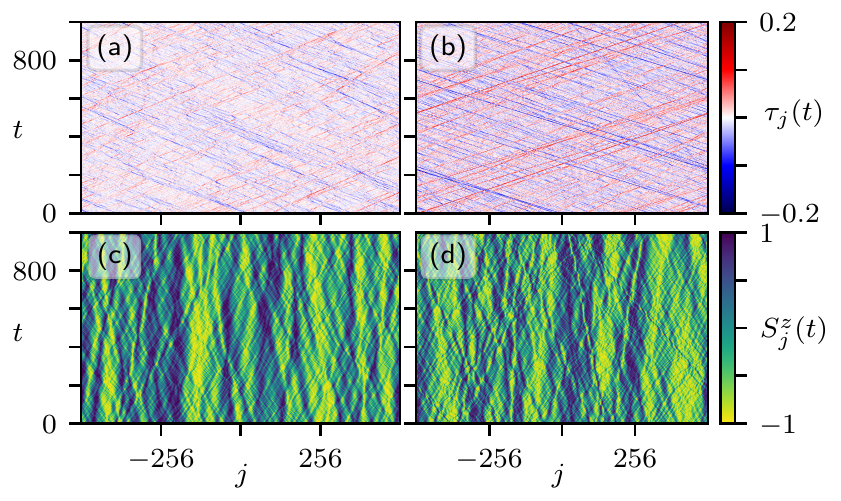}
    \caption{Solitons in the thermal state at $T = 0.1 J$ for the Heisenberg (left) and Ishimori (right) chains. Upper and lower panels show  torsion $\tau_j(t)$, and $S^z_j(t)$ respectively. The ballistic trajectories in the torsion indicate the presence of long-lived solitons in the Heisenberg chain. These ballistic trajectories can also be seen in the plots of $S^z$, where the magnetisation carried by a soliton changes as it moves through the chain -- the mechanism preventing ballistic spin transport.
    }
    \label{fig:thermal_solitons}
\end{figure}

We find that the torsion (\ref{tau}) allows us to track the trajectories of solitons through a thermal background: Figs.~\ref{fig:thermal_solitons}(a) and \ref{fig:thermal_solitons}(b) show the spacetime profile of the torsion $\tau(j, t)$ for a low-temperature thermal state of both the Heisenberg and Ishimori chains. The expected ballistic trajectories of the solitons are clearly observed in the Ishimori chain. Remarkably, very long-lived ballistic trajectories are also observed in the Heisenberg chain. These trajectories can also be seen in the $z$-spin component $S^z(j, t)$ (Figs.~\ref{fig:thermal_solitons}(c) and \ref{fig:thermal_solitons}(d)) -- though, since the magnetisation changes as they propagate, spin is not transported ballistically. 
In a complementary approach, for both the Ishimori and the Heisenberg chain the thermal solitons can also be isolated by surrounding an initial thermal state with a fully-polarised state $\bS = \hat{\boldsymbol{z}}$, and allowing the thermal state to expand into this vacuum during the subsequent dynamics(see Fig.~\ref{fig:inverse_scattering}, \cite{supplemental}).

Having established the existence and nature of the almost integrable behaviour of the Heisenberg chain, we now address the observed KPZ scaling \cite{mcroberts2022anomalous}. In the Ishimori chain, KPZ -- rather than ballistic -- spin transport emerges as follows. As smaller, faster solitons move through larger, slower solitons, they rotate to the `local vacuum'  within the larger soliton. Thus, in any thermal state the magnetisation carried by the smaller solitons is screened on a timsecale set by the rate at which they encounter larger solitons. This argument is qualitatively the same as that for KPZ scaling in the quantum $S = 1/2$ Heisenberg chain that appears in \cite{gopalakrishnan2019kinetic,de2020superdiffusion,DeNardis_2021,Ilievski_2021}. Since the same behaviour is present in the Heisenberg chain at low temperature, this provides a qualitative picture of and explanation for the KPZ regime in spin transport.%

\sectionn{Conclusions}
Our work clearly establishes the existence of a family of solitons in the non-integrable ferromagnetic Heisenberg chain in terms of those known to exist in the integrable Ishimori chain.  
Furthermore, these solitons are shown to exist and be relevant for the dynamics of thermal low-temperature states of the Heisenberg model.
This then allows us to explain the observation of KPZ scaling as a direct consequence of the nearly integrable scattering behaviour of long-lived solitons and the screening of magnetisation during collisions.

The fact that these solitons actually survive and determine the hydrodynamic behaviour at finite temperatures, where correlation lengths are only a few lattice sites, seems truly remarkable, in particular considering that, in that case, the \textit{adiabatically stable solitons are larger than the correlation lengths}. The crossover from this situation to the increasingly normal diffusive transport at higher temperatures is an obvious object for future studies. 

Our work contributes to the broader study of the role of approximate integrability in many-body systems, see, e.g.,\ \cite{Lange_2017}. For this, the Heisenberg chain  provides a suitable setting as, besides the proximity to the Ishimori chain exploited here, it can also be thought of a lattice version of the integrable continuum Landau-Lifshitz model, 
via which route a family of approximate mobile solitons can be obtained by discretisation \cite{Schmidt_2011}, while
in the limit of low temperatures, magnon-type excitations exhibit the usual `emergent integrability' of weakly interacting quasiparticles.  
Our work in particular raises the question about a wider applicability of these ideas about anomalous transport to classical spin models with $SO(3)$ spin-symmetry. It certainly illustrates the point that, even in the very simplest settings, many-body dynamics  still holds many surprises awaiting discovery.

\begin{acknowledgments}
This work was in part supported by the Deutsche Forschungsgemeinschaft under grants SFB 1143 (project-id 247310070) and the cluster of excellence ct.qmat (EXC 2147, project-id 390858490). 
\end{acknowledgments}



\bibliography{refs} 


\onecolumngrid


  \cleardoublepage
  \begin{center}
    \textbf{\large Supplemental Material}
  \end{center}
\setcounter{equation}{0}
\setcounter{figure}{0}
\setcounter{table}{0}
\makeatletter
\renewcommand{\theequation}{S\arabic{equation}}
\renewcommand{\thefigure}{S\arabic{figure}}
\renewcommand{\thetable}{S\arabic{table}}
\setcounter{section}{0}
\renewcommand{\thesection}{S-\Roman{section}}

\setcounter{secnumdepth}{2}

\section{Single soliton solutions of the Ishimori chain \label{Ishimori_solitons}}

We quote here, for ease of reference, the one-soliton solutions of the Ishimori chain \cite{ishimori1982integrable}, and give explicit formulae for some of their physical properties.

The one-soliton solutions are indexed by their inverse-width $R$ and wavenumber $k$. Two further parameters, $x_0$ and $\eta_0$, specify the position of the centre at $t = 0$ and the initial phase of the in-plane spin-components, respectively. The explicit solutions are

\eqa{
S^x_i(t) &= \frac{\sinh2R}{\cosh2R - \cos2k}\sech\xi_{i+1} \left(\cos\eta_i\;(\cosh2R + \sinh2R\tanh\xi_i) - \cos(2k - \eta_i)\right), \nn \\
S^y_i(t) &= \frac{\sinh2R}{\cosh2R - \cos2k}\sech\xi_{i+1} \left(-\sin\eta_i\;(\cosh2R + \sinh2R\tanh\xi_i) - \sin(2k - \eta_i)\right), \nn \\
S^z_i(t) &= 1 - \frac{\sinh^2 2R}{\cosh2R - \cos2k}\sech\xi_i\sech\xi_{i+1},
\label{sol_exact}
}
where
\eqa{
\xi_n(t) &= 2R\left(n - x_0 - \frac{1}{2}\right) - 2t\sinh2R \sin2k, \nn \\
\eta_n(t) &= -2k\left(n - x_0 - \frac{1}{2}\right) + \eta_0 + 2t\left(1 - \cosh2R\cos2k \right).
}

These solutions obey a lattice version of the travelling wave ansatz,
\eqa{
S_{i+n}^x(t + n/v) &= \cos(n\omega/v) S_i^x(t) - \sin(n\omega/v) S_i^y(t), \nn \\
S_{i+n}^y(t + n/v) &= \sin(n\omega/v) S_i^x(t) + \cos(n\omega/v) S_i^y(t), \nn \\
S^z_{i+n}(t + n/v) &= S^z_i(t),
}
where $v$ is the velocity of the soliton, and $\omega$ is the internal frequency, which, in terms of the physical parameters $R$ and $k$, are given by
\eqa{
v = \frac{\sinh2R \sin2k}{R},\;\;\;\;  \omega = 2 k v + 2(\cosh2R\cos2k - 1).
\label{freq_and_velocity}
}

The limit $v\rightarrow 0$ (i.e., $k\rightarrow0$) implies that the $z$-components become stationary, and the in-plane components precess with the internal frequency $\omega$.

The total energy (measured by the Heisenberg Hamiltonian (\ref{H_Heisenberg})), magnetisation (defined as the difference from the vacuum state, i.e., $M = \sum_{n\in\mathbb{Z}} (1 - S_n^z)$), and torsion carried by a soliton are, respectively,

\eqa{
E = 4\tanh2R, \;\;\;\; M = \frac{2\sinh 2R}{\cosh2R - \cos2k}, \;\;\;\; \tau = 2Rv.
\label{EMT}
}

\section{Single stationary solitons in the Heisenberg chain}

In this section, we construct exact, non-dissipative (i.e., soliton) solutions of the classical Heisenberg chain. The equations of motion are
\eqn{
\dot{\bS_i} = J \bS_i \times \left( \bS_{i-1} + \bS_{i+1} \right).
}

It will be most expedient to write the equations of motion in the canonical $(\phi, z)$ co-ordinates, where they take the form
\eqn{
\dot{z_i} = J\sqrt{1 - z_i^2} \; \biggl[\sqrt{1 - z_{i+1}^2} \sin(\phi_{i+1} - \phi_i)
- \sqrt{1 - z_{i-1}^2} \sin(\phi_{i} - \phi_{i-1})\biggr],
\label{z_eq_H}
}
\eqn{
\dot{\phi_i} = J \frac{z_i}{\sqrt{1 - z_i^2}} \left[\sqrt{1 - z_{i+1}^2} \cos(\phi_{i+1} - \phi_i)
+ \sqrt{1 - z_{i-1}^2} \cos(\phi_{i} - \phi_{i-1}) \right]
-J (z_{i+1} + z_{i-1}).
\label{phi_eq_H}
}

As discussed in the main text, we use an ansatz based on the structure of the stationary ($k = 0$) Ishimori solitons. We set $z_0 = -1$, $\phi_{i<0}(0) = \varphi$, and $\phi_{i>0}(0) = \varphi + \pi$, for some arbitrary constant $\varphi$. If $\phi_i$ precesses with a uniform frequency, $\phi_i(t) = \phi_i(0) + \omega t$, then $\dot{z}_i = 0$ $\forall i$ (the sine factors vanish at all times, unless they contain $i = 0$, in which case one of the square root factors vanishes instead).

We thus aim to find a set of $z_i$ s.t. $\phi_i(t) = \phi_i(0) + \omega t$, for some chosen uniform frequency $\omega$ which characterises the soliton. Inserting these conditions into Eq.~(\ref{phi_eq_H}), we obtain the consistency equations
\eqn{
\omega = J \frac{z_i}{\sqrt{1 - z_i^2}} \left(\sqrt{1 - z_{i+1}^2} + \sqrt{1 - z_{i-1}^2} \right)
-J \left(z_{i+1} + z_{i-1}\right),
}
which, rearranged for $z_i$, become
\eqn{
\frac{z_i}{\sqrt{1 - z_i^2}} = \cot(\theta_i) = \frac{\omega/J + z_{i+1} + z_{i-1}}{\sqrt{1 - z_{i+1}^2} + \sqrt{1 - z_{i-1}^2}}.
\label{eq:SI_sol_H}
}

We need $z_{-1} = z_1$ to ensure the exchange field at $i = 0$ is parallel to $\hat{\boldsymbol{z}}$, which implies $z_{-i} = z_i$ $\forall i$. It thus suffices to solve the consistency equation for $i > 0$. We cannot do this in closed form, but the required $z_i$ may be obtained numerically to arbitrary precision, for any choice of $\omega$.

We note that $\theta: (0, \pi) \rightarrow \mathbb{R}$, $\theta \mapsto \cot(\theta)$ is bijective, so, given $z_{i-1}$ and $z_{i+1}$, Eq.~(\ref{eq:SI_sol_H}) can be inverted for a unique $z_i$ (so long as at least one of $z_{i-1}$, $z_{i+1}$ is not at the poles). 
To solve the consistency equations iteratively, we choose the frequency $\omega$, and begin with the sequence $z_2 = 0$, $z_{i>2} = 1$ (though all that is required is that $-1 < z_2 < 1$). We then obtain $z_1$ from Eq.~(\ref{eq:SI_sol_H}). In the second step, we first solve for a new $z_2$, and then re-solve for $z_1$. Continuing the pattern, at the $n^{th}$ step we solve for $z_n$, and then sweep back to $z_1$. 

Of course, only finite-size solitons can be constructed numerically. We carry out the above procedure until we reach some final $z_N$ (effectively, we approximate $z_{i>N} = +1$, and the chain that this describes has $2N + 1$ sites). To improve the solution, we then perform a number of sweeps in the forward direction, starting with $z_1$ and solving up to $z_N$. 

To measure how well the numerical solution solves the consistency equations, we define the cost function 
\eqn{
\mathcal{C}(\{z_i\}) = \left( \sum_{i=1}^{N+1} \left|\; z_i\left(\sqrt{1 - z_{i+1}^2} + \sqrt{1 - z_{i-1}^2}\right) - \sqrt{1 - z_i^2}\biggl(\omega/J + z_{i+1} + z_{i-1}\biggr)\; \right|^2 \right)^{1/2},
}

\begin{figure}[!htb]
    \centering
    \includegraphics{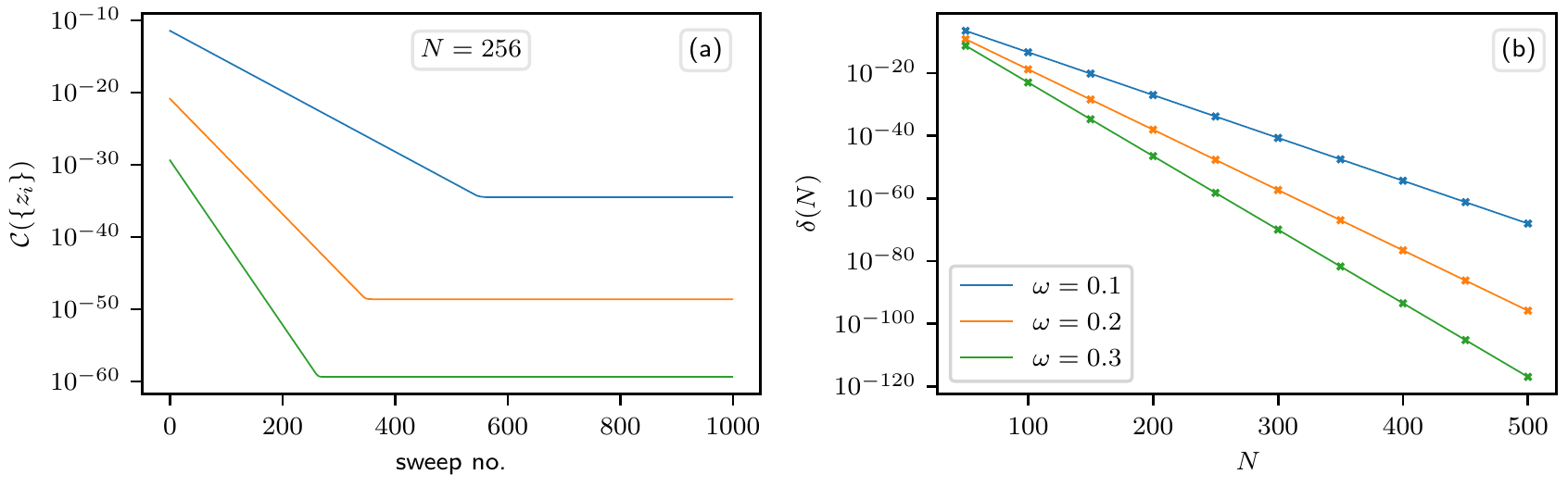}
    \caption{Convergence of the iterative procedure for solving the consistency equations (\ref{eq:SI_sol_H}) for selected frequencies and finite sizes. (a) shows the exponential convergence of the cost function to the plateau value, $\delta(N)$, with the number of forward sweeps ($z_1$ to $z_N$) -- the starting values are those obtained after all the backward sweeps have been completed. (b) shows that $\delta(N) \rightarrow 0$ exponentially as $N \rightarrow \infty$.}
    \label{fig:convergence}
\end{figure}

shown in Fig.~\ref{fig:convergence} versus the number of sweeps performed for select frequencies. We observe that the solution converges exponentially with the number of sweeps, to a plateau value $\delta(N)$ that exponentially decreases with system size.

Whilst we are unable to rigorously prove that the cost function converges to zero, we conjecture, based on the numerical solution, that the iterative procedure defines an exact solution of the consistency equations in the limit $N \rightarrow \infty$, and, thus, a stationary soliton of the Heisenberg chain.

\section{Adiabatic stability of the solitons}

In this section we analyse the stability of the adiabatically constructed solitons in the Heisenberg chain. As mentioned in the main text, we use a soliton solution of the Ishimori chain as the initial conditions, and perform numerical time evolution whilst continuously tuning the model from the Ishimori chain to the Heisenberg chain. 

The adiabatic time used to calculate the existence diagram (Fig.~\ref{fig:solitons}) was $t_A = 10^5 J^{-1}$, on a chain of $L = 1024$ sites and periodic boundary conditions (PBCs). The final time of the simulation was $t_f = 2 t_A$, i.e., the state was evolved with the Heisenberg equations of motion for a further time of $10^5 J^{-1}$ after the completion of the adiabatic process.

We use the following criteria to assess whether the resultant state is a solitary wave solution of the Heisenberg chain:
\begin{itemize}
    \item $\forall t$ there is a unique local minimum of $z_i$, up to a tolerance of $10^{-8}$.
    \item $\forall t > t_A$, the unique local minimum propagates with a constant velocity $v$. More precisely, since the centre can only be measured to integer precision, the condition is: $\exists\; v, x_0 \in \mathbb{R}$ s.t. $\forall t > t_A$, $|\mathrm{min}_i z_i - (x_0 + v t)| < 1$.
    \item $\forall t > t_A$, the torsion $\tau = \sum_i \tau_i$ is constant in time, up to a tolerance of $10^{-8}$. The torsion is not a conserved quantity of the Heisenberg chain, but is constant for the solitary wave solutions.
\end{itemize}

The first condition ensures that no pulses are emitted, as happens if a soliton of the Ishimori/Heisenberg chain is evolved with the other Hamiltonian without an initial adiabatic interpolation; the second condition ensures the soliton propagates with a constant velocity; and the third ensures that the soliton is not decaying by slowly spreading out in space.

The numerical values of the tolerances are somewhat arbitrary, but they are necessary to account for numerical error and finite-size effects (the soliton solutions, even in the Ishimori chain, are only exact in the limit $L \rightarrow \infty$).

\begin{figure}[!htb]
    \centering
    \includegraphics{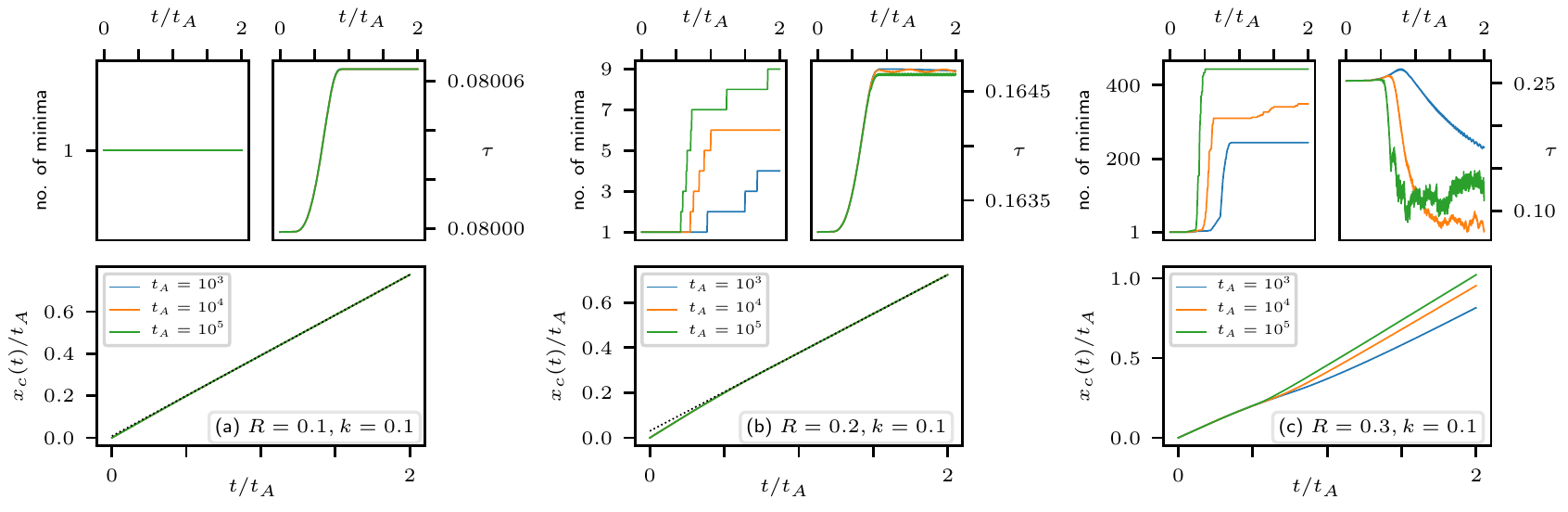}
    \caption{Adiabatic stability of the solitons. (a) shows that the Ishimori soliton $R = 0.1, k = 0.1$ is adiabatically connected to a Heisenberg soliton: no additional local minima are generated, the centre propagates with a constant velocity after $t = t_A$, and the profile retains its new shape, as evidenced by the fact that the torsion $\tau$ is constant for $t > t_A$. Moreover, these conclusions hold for all three values of $t_A$. (b) and (c) show two Ishimori solitons that are not adiabatically connected to Heisenberg solitons, since additional local minima are generated in both cases, and the torsion is not constant (though (b) is very close to the edge of the existence diagram, and so is much more stable than (c)). Increasing $t_A$ by a factor of $100$ does not improve the stability of the soliton.}
    \label{fig:stability}
\end{figure}

\section{Soliton-soliton scattering}
As discussed in the main text, when two Ishimori solitons collide, their asymptotic trajectories are unchanged, apart from a `phase-shift' $\Delta(R, k; R', k')$. The phase-shift directly corresponds to a shift of the trajectory of the soliton by that number of sites: the centre of a soliton after a sequence of collisions is $x_c(t) = x_c(0) + v t + \sum_{\lambda} \Delta(R, k; R_{\lambda}, k_{\lambda})$.

\begin{figure}[!htb]
    \centering
    \includegraphics{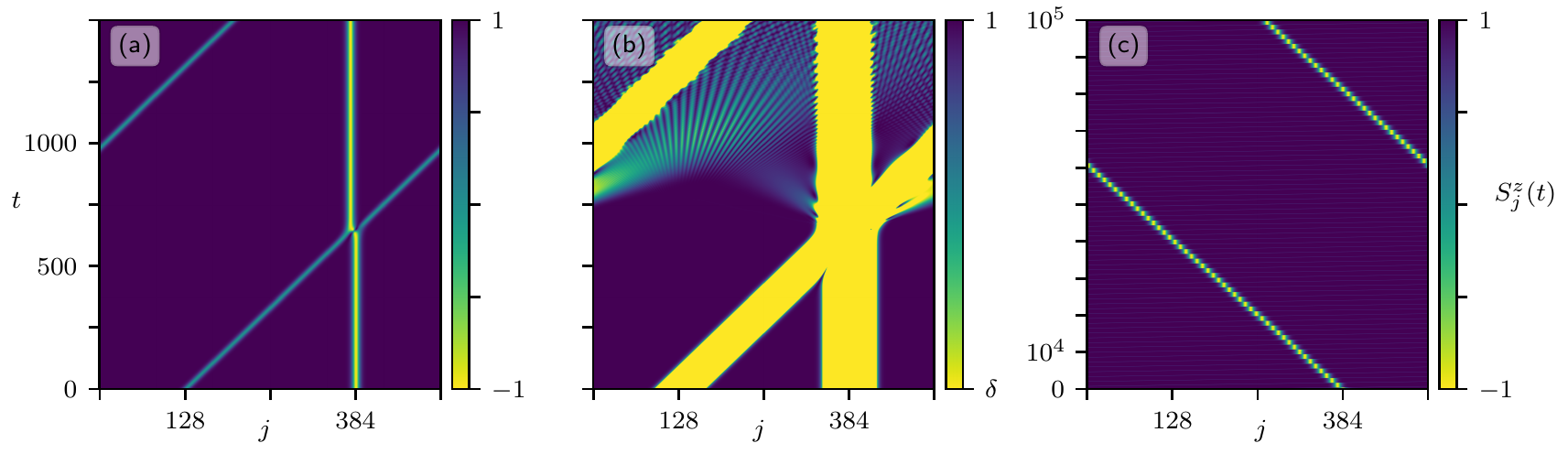}
    \caption{
    ($\delta = 1 - 10^{-6}$). Colour-scales show the $z$-components. Almost integrable scattering between solitons $(0.1, 0.1)$ and $(0.1, 0)$ in the Heisenberg chain. (a) shows the scattering event -- the phase-shift of the stationary soliton is clearly visible. (b) shows the same scattering event, but the colour-scale is focused on small deviations from the vacuum $S^z = +1$, which shows that the collision does emit some small amount of radiation. (c) shows approximately $80$ scattering events with the same solitons ($L = 512$, PBCs), with the same phase-shift on every collision -- the slowly moving bright streak is the \textit{stationary} soliton being shifted upon collisions with the moving soliton.}
    \label{fig:scatter1}
\end{figure}

Even though the Heisenberg chain is not integrable, this picture remains true to a surprising extent. We show in Fig.~\ref{fig:scatter1}(a) the scattering of a soliton $(R_1 = 0.1, k_1 = 0.1)$ incident upon a stationary soliton $(R_2 = 0.1, k_2 = 0)$, showing a clear shift after the collision and unchanged asymptotic velocities of the solitons. However,   Fig.~\ref{fig:scatter1}(b) shows that there is a small amount of radiation emitted during this event by focusing the colour-scale on small deviations from $S^z = +1$. This does not, however, immediately destabilise the solitons -- in Fig.~\ref{fig:scatter1}(c), the same solitons (under PBCs with $L = 512$) exhibit no signs of decay after c. $80$ scattering events.

\begin{figure}[!htb]
    \centering
    \includegraphics{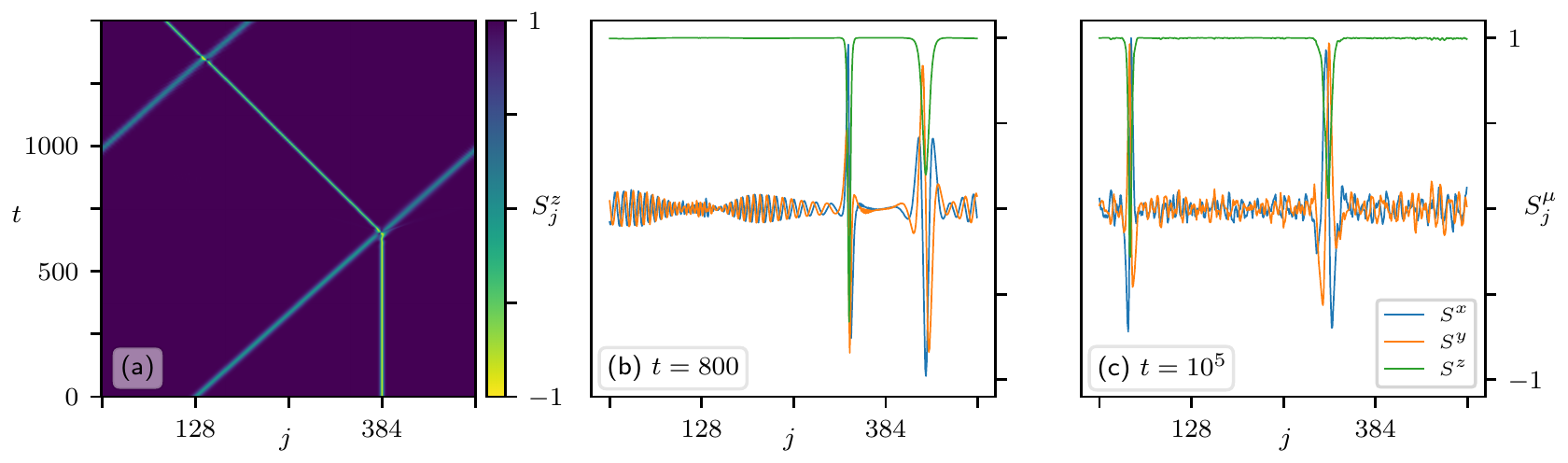}
    \caption{A destructive scattering event between solitons with parameters $(0.1, 0.1)$ and $(0.25, 0)$. The colour-scale in (a) shows the $z$-component. The first scattering event in (a) shows that the initially stationary soliton acquires a velocity, although the moving soliton is relatively well-preserved; at the second scattering event, the two scattered solitons appear to pass through each other without incident. (b) shows the scattered solitons and the decaying pulses emitted from the first collision -- on an infinite chain, the pulses would dissipate, but at finite size they become noise. (c) shows the solitons at a much later time, $t = 10^5$ -- despite both circling round the chain approximately 100 times, colliding on each pass, and propagating in the noise generated by the first collision, they are remarkably well-preserved.} 
    \label{fig:scatter2}
\end{figure}

To this point we have discussed the scattering in the context of the similarity to the integrable model. We should point out, however, the possibility of more destructive scattering events. We show in Fig.~\ref{fig:scatter2} the collision of solitons $(R_1 = 0.1, k_1 = 0.1)$ and $(R_2 = 0.25, k_2 = 0)$. This collision releases a significant amount of radiation. This radiation appears to be dissipative, in the sense that it spreads out over the chain with no persistent localised features. 

However, whilst the incident solitons are somewhat changed by the collision, there are clearly two solitons which emerge from the vertex. These modified solitons are then apparently stable to further collisions with each other (using PBCs), surviving c. $100$ more scattering events, with no sign of further decay, even up to $t = 10^5$ (Fig.~\ref{fig:scatter2}(c)).

Notably, this also indicates that these solitons are stable to (sufficiently) weak fluctuations of the background, suggesting they should remain stable in low-temperature thermal states.

\subsection{Phase-Shifts}
We now address the question of the scattering phase-shifts $\Delta$. In Fig.~\ref{fig:scattering}(d) in the main text we show the phase-shifts experienced by a soliton $(R, k = 0)$ upon collision with a $(0.1, 0.1)$ soliton. In the Ishimori chain, these four parameters ($R, k, R', k'$) completely determine the phase-shift. This does not appear to be the case in the Heisenberg chain, where the phase-shift appears to fluctuate for different scattering events -- as a result of small differences in the initial conditions of the solitons. Though we note that this dependence vanishes for sufficiently large solitons. 

\begin{figure}[!htb]
    \centering
    \includegraphics{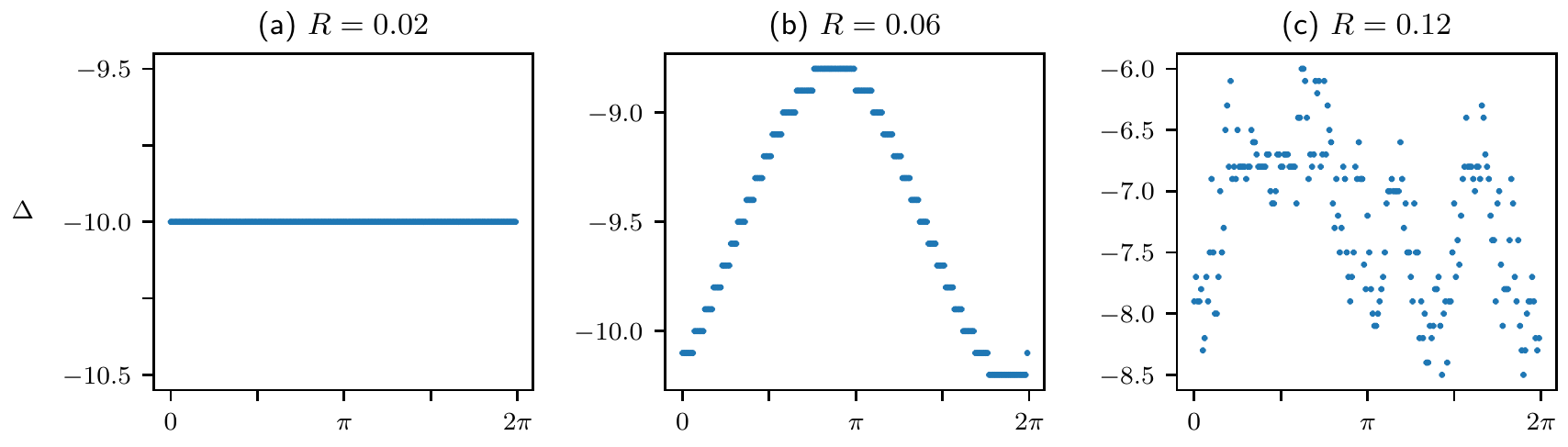}
    \caption{Dependence of the scattering phase-shift $\Delta(R, 0; 0.1, 0.1)$ on the phase difference of the colliding solitons, for different, illustrative values of $R$. (The actual value of the relative phase is not measured, so the starting point is arbitrary -- but a full $2\pi$-sweep is mapped out). In the integrable case, there would be no dependence on $\phi_0$. In the Heisenberg case, this remains true for large solitons ($R = 0.02$), but then gives way to a periodic dependence ($R = 0.06$), and finally becomes effectively random as the target soliton becomes narrower ($R = 0.12$). In all cases, however, the soliton never acquires any velocity from the collision, as otherwise $\Delta$ would not be measurable.}
    \label{fig:scatter3}
\end{figure}

In Fig.~\ref{fig:scatter3} we examine the dependence of the phase-shift on the overall phase of the target soliton -- that is, we uniformly rotate the target soliton about the $z$-axis through some angle $\phi_0$ (the incident soliton is not rotated, thus changing the relative phase). Since we can only measure the position of the centre of the soliton to an accuracy of one site, the phase-shifts are calculated by averaging over ten scattering events with the same incident soliton -- giving a resolution of $0.1$ sites. 

We observe in Fig.~\ref{fig:scatter3}(a) that, for sufficiently large target solitons, the phase-shift exhibits no dependence on $\phi_0$. In Fig.~\ref{fig:scatter3}(b) at intermediate widths there is a smooth, periodic variation of the phase-shift with $\phi_0$. Fig.~\ref{fig:scatter3}(c), however, shows that for narrower solitons the phase-shift becomes highly sensitive to the initial conditions determined by the angle $\phi_0$.

\section{`Inverse scattering' simulations}

In the main text we observed the presence of solitons in thermal states of the Heisenberg chain via the torsion, which displays clear, long-lived ballistic trajectories within the chain at a given temperature $T$. There is an interesting complementary picture, in the spirit of the inverse-scattering transform \cite{doyon2019lecture}, where we take a thermal region of length $N$ and immerse it in a fully-polarised state $\bS = \hat{\boldsymbol{z}}$ of length $L >> N$.

In the integrable case, the thermal state expands into the surrounding quasiparticle (soliton) vacuum, and the solitons, since they have different velocities, become separated in space. If $L \rightarrow \infty$ (before $N \rightarrow \infty$), this permits a description of the initial thermal state on $N$ sites in terms of the asymptotic trajectories of the solitons. In particular, this is possible since scattering in the integrable case is not \textit{dissipative} and does not change the soliton velocities -- thus, the long-time state of well separated solitons is guaranteed to contain the very same solitons as the initial state.

\begin{figure}[!htb]
    \centering
    \includegraphics{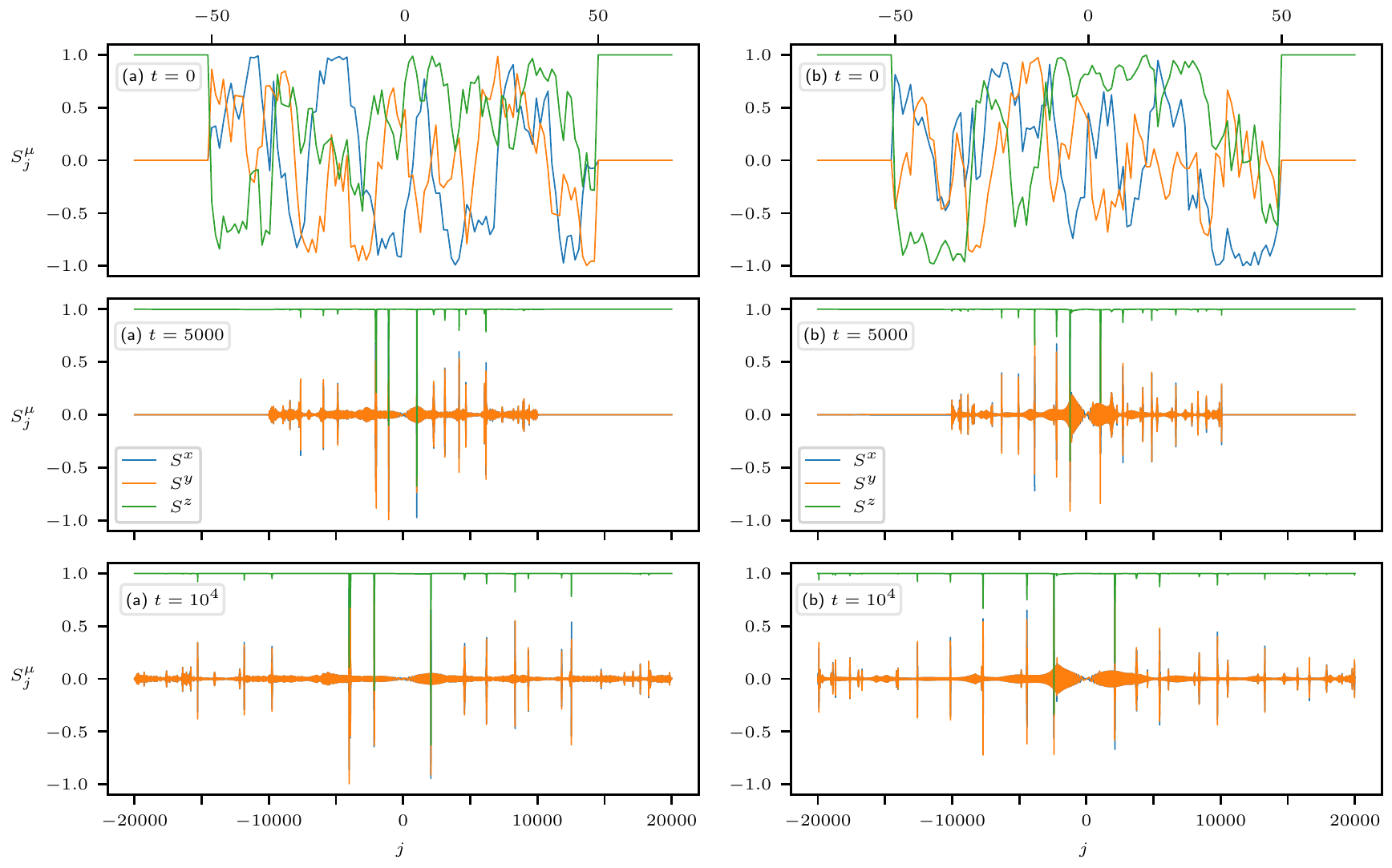}
    \caption{Examples of the expansion of a thermal state connected to a fully polarised state (the `inverse-scattering experiments'). The initial configurations are thermal states on a chain of length $N = 100$ at $T = 0.1$. They are then connected to a fully polarised state -- the total length of the system is $L = 4\times10^4$. (a) evolves under the Heisenberg chain dynamics; (b) evolves under the Ishimori chain dynamics. 
    The expected inverse-scattering is observed in the Ishimori chain, where the initial thermal region expands into a dilute gas of well-separated solitons. Remarkably, the picture in the Heisenberg chain is, at least qualitatively, strikingly similar. The principal difference is that the Ishimori chain has solitons that are both narrow and fast, cf. the existence diagram (Fig.~\ref{fig:solitons}(a))}
    \label{fig:inverse_scattering}
\end{figure}

Remarkably, in Fig.~\ref{fig:inverse_scattering} we observe qualitatively very similiar dynamics for the non-integrable Heisenberg chain. Specifically, it appears that during expansion from a thermal state, well-defined spatially localised solitonic excitations emerge that propagate 
non-dissipatively, at least on numerically accessible length- and timescales, allowing them to become well separated in space.

\end{document}